\documentclass[dvips]{arxstspdf}
\usepackage{flushend}

\volume{25}
\issue{2}
\pubyear{2010}
\firstpage{166}
\lastpage{169}
\doi{10.1214/10-STS308C}
\referstodoi{10.1214/09-STS308}

\begin{document}
\begin{frontmatter}

\vspace*{8pt}
\title{Comment: How Should Indirect Evidence Be Used?}
\runtitle{Comment}

\begin{aug}
\author{\fnms{Robert E.} \snm{Kass}\ead[label=e1]{kass@stat.cmu.edu}}
\runauthor{R. E. Kass}

\affiliation{Carnegie Mellon
  University}

\address{Robert E. Kass is Professor,
Department of Statistics, Center
  for the Neural Basis of
  Cognition, and Machine Learning Department, Carnegie Mellon
  University, Pittsburgh, Pennsylvania, 15217 USA \printead{e1}.
}

\end{aug}

\begin{abstract}
Indirect evidence is crucial for successful statistical practice.
Sometimes, however, it is better used informally.
Future efforts should be directed toward understanding better the
connection between statistical methods and scientific problems.
\end{abstract}


\end{frontmatter}

When Brad Efron speaks about statistical theory and methods we should
pay attention.
In his talk, as he prefers to call it, he returns to
a theme that has surfaced in previous ruminations: his unease with the
foundations of statistics and his feeling that there is something
missing.  In this version he highlights indirect evidence as the
aspect of statistical reasoning in need of the theory he yearns for.

The framework of statistical decision theory
was created over 50 years ago for small, well-defined problems.
Efron seeks an extension
to accommodate large datasets where individual observations bear an
uncertain relationship to one another.  He seems to think such an
extension is possible and important for the future of the discipline.
Perhaps he is right but, I'm sorry to say, I don't get it.
In trying to understand the role of indirect
evidence I would examine not theoretical
foundations but, instead,
the relationship of statistical
methodology to scientific inference in the context
of specific applications.

Efron begins by citing clinical trials as furnishing ``direct evidence''
about a question of interest. It is easy to see what he means, but
the stereotypical problem in a clinical trial is
somewhat special because all the relevant
background knowledge has been focused on producing a simple treatment
comparison, a comparison that statistical inference will evaluate in a
final declarative step.  Clinical trials are aimed at treatment
policy, so decision theory is highly relevant. In particular, the
concepts of type I and type II error have an unusual immediacy because
decisions about patients must be made across a large population.

In the scientific applications I am familiar with, statistical
inferences are important, even crucial, but they constitute
intermediate steps in a chain of inferences, and they are relatively
crude. As Jeffreys pointed out long ago, inferences may be based on
estimates and standard errors, and they typically need to be accurate
only to first order.  Similarly, in using the bootstrap we can get by
with a fairly small number of observations from the bootstrap
distribution because simulation uncertainty quickly becomes smaller
than statistical uncertainty. Furthermore, statistical uncertainty is
typically smaller than the unquantified aggregate of the many other
uncertainties in a scientific investigation. I~tell my students in
neurobiology that in claiming
statistical significance I get nervous unless
the $p$-value is much smaller than 0.01,
and if some refinement of an estimate or $p$-value changes a
conclusion, that indeterminacy itself becomes the story. To be
convincing, the science needs solid statistical results, but
in the end only a qualitative summary is likely to
survive.  For instance, in Olson et al.  (\citeyear{OlsonEtAl2000}), my first
publication \mbox{involving} analysis of neural data, more than a dozen
different statistical analyses---some of them pretty meticulous,
involving both bootstrap and MCMC---were reduced to the main message
that among 84 neurons recorded from the supplementary eye field,
``Activity reflecting the direction of the [eye movement] developed
more rapidly following spatial than following pattern cues.''  The
statistical details reported in the paper were important to the
process, but not for the formulation of the basic finding.  Such
settings seem to me vastly different than that conceptualized by
decision theory. In judging the role of statistical analysis within
the general scientific enterprise, I prefer Fisher and Jeffreys to
Neyman and Savage.

If science is such a loose and messy process, and inferences so rough
and approximate, where does all the statistical effort go?  In my
view, Jeffreys got it right. State-of-the-art analyses may take
months, but they usually come down to estimates and standard errors.
The biggest news in the early 1990s was the development, understanding, and
propagation of MCMC, which has had an enormous influence on
statistical practice.  The ``Bayesian revolution,'' however, in my
view, is a misnomer. The most important method in Bayesian inference
is what Fisher called the method of maximum likelihood. Most of the
time what
those people running Markov chains are doing is, essentially,
computing MLEs. The ``revolution'' is really a maximum
likelihood/Bayesian synthesis based on EM and Gibbs sampling, and
their generalizations. It has shown the power of the
insights articulated by Fisher and Jeffreys.
(With only a bit of a stretch
Dirichlet processes and their relatives may be included
as extensions of the basic ideas.)
What has advanced
over the years is the complexity of the problems we are able to
attack, not the fundamental framework.

Data analytic methods comprise both data\break manipu\-lation---including
estimates and standard\break errors---and interpretation. Manipulation
involves\break the mechanics of statistical inference, interpretation its
logic. If I am reading him correctly,
Efron seems to be concerned primarily with the latter.  To
exemplify the kind of ``difficult new problems'' he has in mind Efron
uses a hypothetical issue in applying FDR to neuroimaging, half-brain
versus whole-brain analysis. When fMRI first hit the scene, almost 20
years ago, a statistician told me of psychologists who were doing
many thousands of voxel-wise $t$-tests simultaneously. The standard
method was
to line up the test statistics in ascending order of magnitude, or
descending order of $p$-value, and to pick a threshold that gave them
suitable results. In our statistician's n\"aivety, we shook our heads
with indignation. (I was
so much older then$\,\ldots.$)  Then FDR came along and provided precisely
the same method of data manipulation, but furnished a new interpretation.
And it is a wonderful interpretation, very helpful. I think
we all appreciate it.
However, as its chief accomplishment is to bless the procedure
psychologists were already using (but feeling uncomfortable about, due
to problems in controlling family-wise error rate), it is hardly
surprising that they like it.  I am not by any means an expert in
neuroimaging, let alone in diffusion tensor imaging, but I am dubious
about the scientific importance of half-brain versus whole brain FDR.
I would guess the bigger issues involve connectivity across voxels and
the hazards of warping brains from different individuals algorithmically so
that their voxels are aligned. I should think a more pressing problem
would be to devise within-subject expressions of uncertainty about
white matter fibers in regions of potential interest, and a method of
combining such things across subjects, within groups. (Apparently
initial steps in getting local DTI uncertainy have been taken by
Zhu et al., \citeyear{ZhuEtAl2007}, and by
Efron's former student Armin Schwartzman, \citeyear{Schwartzman2007}, whom he cites.)

In picking on this example I should acknowledge that everyone who
discusses statistical methods per se abstracts away from details
of the scientific\break problem---Fisher and Jeffreys did so, too, and it is
unavoidable. I~just do not yet understand the logical difficulty Efron
is concerned about.
While I certainly agree that the use of indirect
evidence is a major challenge, especially in dealing with large datasets, it seems to me that with the passage of time our existing
logical frameworks are
treating us remarkably well. Nor do I see any problem with being
Bayesian in one analysis and frequentist in another, or even
combining the two in a single swoop. The heyday
of decision theory referenced by Efron occurred during a time
that emphasized pure theory in many parts of academic life.
Now we are in a much more utilitarian period and many of us
are content to use whatever seems best suited for the task
in front of us. As I have argued
elsewhere (Kass, \citeyear{Kass2010}), I~believe a straightforward
philosophy I have called \textit{statistical pragmatism}
can incorporate both Bayesian and frequentist inference.

It is tempting
to try to formalize the many aspects of direct and
indirect evidence that must get
weighed together, and it is possible to do so\break Bayesianly.
Like Efron, however, I am wary.
In Kass (\citeyear{Kass1983}) I commented on a very nice, but ambitious
paper by
DuMouchel and Harris in which they used a Bayesian hierarchical model
to combine evidence about cancer across species:

\begin{quotation}
The Bayesian approach has its difficulties, for while it is surely
desirable to express [knowledge] explicitly, in particular through models,
it is often difficult to do so accurately. Lurking beside each
analysis are the interrelated dangers of oversimplification,
overstated precision, and neglect of beliefs other than the analyst's.
\end{quotation}

Where I may disagree with Efron is that I do not think it is
likely to be fruiful to try some other formalization. The problem
in such situations
is not inadequacy of logic, but rather the unclear relevance
of the related evidence. As I said in Kass (\citeyear{Kass1983}), I would not
want to apply formal methods in the absence of pretty solid
theoretical or empirical knowledge.

In tackling the complexities of real-life science, real-life clinical
trials, or real-life policy decisions,
statisticians can bring unique insight based on
statistical expertise combined with nontrivial experience
in the substantive area. They then exercise good sense as they
go along. My statistical bioinformatics colleague
Kathryn Roeder put this well recently
when she told me, ``I violate type I error all
the time. And do you know why? I actually want \textit{to find} those
genes!''
As Emery Brown and I emphasized in a recent article (Brown and Kass, \citeyear{BrownKass2009}), this requires new attitudes about training. It also
requires an altered notion
of our relationship to our collaborators: as Brown and I said,
we should put to rest their characterization (used here
by Efron) as ``clients'' and, instead, agree to share responsibility
for all aspects of scientific inference---not just statistical ones.
In attempting to understand the anatomical basis of dyslexia,
of course it matters which part of the
brain we focus on, but the choice
can not be made in terms of abstract statistical arguments.
It should result
from closely-knit
statistical, neuroimaging, neuroanatomical,
and psychological judgment.

Now, I am pretty confident that Efron will
agree about this. I bring it up because we
judge statistical methods by the two rather
different standards of
theoretical performance (evaluated either by mathematics or
by simulation studies) and apparent effectiveness in
answering an applied question. I find it impossible to think about
either one without considering the other, and failure on either front
serves to veto further contemplation.

I understand Efron's ``indirect evidence'' to include anything that
could, in principle, be used to help formulate a prior for a Bayesian
analysis. My impulse is to come at indirect evidence from an applied
perspective, and I think an uneasiness much like Efron's motivated me
in 1990 to begin organizing the workshop series \textit{Case Studies in
  Bayesian Statistics}. I had the lofty goal of identifying and
describing key steps in using scientific and technological knowledge
to build good Bayesian models and priors, so as to help turn the art
of Bayesian statistical practice into a science.  The idea was to gain
understanding of statistical effectiveness by examining methods
carefully in an applied context, and I pointed to Mosteller and Wallace
(\citeyear{MostellerWallace1964}) as the archetype.
However, I must admit that while the
workshops have been very successful as meetings, they never made much
progress on the big agenda. The reason was simply that the audience
was too diverse scientifically, so that speakers could not get very
far into the details of connecting statistics to science that I
originally had in mind.  In 2002 Emery Brown and I began a series of
meetings \textit{Statistical Analysis of Neural Data} which are broader
statistically but, due to their narrower scientific focus, may
actually be more successful in
providing material for learning about statistical methods.

I have been negative about comprehensive\break Bayesian analyses, yet I have
spent much time and effort trying to understand and promote Bayesian
methods. In many circumstances Bayesian methods are great, and very
hard to beat.  The nonparametric regression method BARS, for example
(DiMatteo, Genovese and Kass, \citeyear{DiMatteoGenoveseKass2001}), began with existing frequentist
and Bayesian results on free-knot splines and used reversible-jump MCMC
to great advantage; it was difficult to code properly and takes a long
time to run on even modestly sized datasets, but I have not seen
another general method produce smaller mean-squared error and more accurate
coverage probabilities, and I would be surprised to find an
alternative that works much better for the problem we designed BARS to
solve, namely Poisson regression with smoothly varying means, which is
suitable for fitting neural firing rate intensity functions.  BARS
illustrates a general truism: we may
expect Bayes to work well if there is solid knowledge about the
problem that can lead to useful formalization, if one is
willing to spend the time it takes to be careful, and if one has the
computing resources to get the job done. These are big ``ifs.''
The challenge of indirect evidence is to figure out when they are
satisfied.

\section*{Acknowledgment}
This work was supported in part by NIH Grant MH064537.

\end{document}